\documentstyle[12pt,epsfig]{article}
\def \b{{\cal B}}
\def \bea{\begin{eqnarray}}
\def \beq{\begin{equation}}
\def \bo{B^0}

\def \cn{Collaboration}
\def \eea{\end{eqnarray}}
\def \eeq{\end{equation}}
\def \ite{{\it et al.}}

\def \ob{\overline{B}^0}
\def \ok{\overline{K}^0}
\def \rt{r_{\tau}}
\def \s{\sqrt{2}}
\begin{document}

\begin{flushright}
TECHNION-PH-2001-35\\
EFI 01-42 \\
hep-ph/0109238 \\
September 2001 \\
\end{flushright}

\renewcommand{\thesection}{\Roman{section}}
\renewcommand{\thetable}{\Roman{table}}
\centerline{\bf IMPLICATIONS OF CP ASYMMETRY LIMITS}
\centerline{\bf FOR $B \to K \pi$ AND $B \to \pi \pi$}
\medskip
\centerline{Michael Gronau}
\centerline{\it Physics Department, Technion -- Israel Institute of Technology}
\centerline{\it 32000 Haifa, Israel}
\medskip
\centerline{Jonathan L. Rosner}
\centerline{\it Enrico Fermi Institute and Department of Physics}
\centerline{\it University of Chicago, Chicago, Illinois 60637}
\bigskip

\begin{quote}

Recent experimental limits for the direct CP asymmetries
in $B^0 \to K^+ \pi^-$, $B^+ \to K^+ \pi^0$, $B^+ \to K^0 \pi^+$, and $B^0 \to
\pi^+ \pi^-$, and for the indirect CP asymmetry in $B^0 \to \pi^+ \pi^-$,
are combined with information on CP-averaged branching ratios to shed light on
weak and strong phases.  At present such bounds favor $\gamma \ge 60^\circ$
at the $1 \sigma$ level.  The prospects for further improvements are discussed.

\end{quote}

\leftline{\qquad PACS codes:  12.15.Hh, 12.15.Ji, 13.25.Hw, 14.40.Nd}

\section{Introduction}

The decays of $B$ mesons to the charmless final states $\pi \pi$ and
$K \pi$ are a rich source of information on the fundamental parameters of
the Cabibbo-Kobayashi-Maskawa (CKM) matrix, but the extraction of this
information from data requires the separation of weak interaction effects
from strong-interaction quantities such as magnitudes of operator matrix
elements and strong phases.  A number of model-independent analyses of these
systems \cite{GRKpi,NR,GRpipi,PP} have shown that when one combines data on
CP asymmetries with branching ratios of CP-averaged final states, one can
separate the strong interaction effects from fundamental CKM parameters,
obtaining useful information on both sets of quantities.

In the present paper we apply several of these analyses \cite{GRKpi,NR,GRpipi}
to the decays $B \to K \pi$ and $B \to \pi \pi$, using new upper limits quoted
by the CLEO \cite{CLEOasy}, BaBar \cite{BaBasy1,BaBasy2}, and Belle
\cite{Belasy} Collaborations for several CP-violating asymmetries in these
decays, as well as updated CP-averaged branching ratios for these states.
Comparison of the CP-averaged rate for $B^0
\to K^+ \pi^-$ with that for $B^+ \to K^0 \pi^+$, given a small strong phase
difference, excludes $31^\circ \le \gamma \le 60^\circ$ for
the weak phase $\gamma \equiv {\rm Arg}(-V^*_{ub}V_{ud}/V^*_{cb}V_{cd})$,
while comparison of $B^+ \to K^+ \pi^0$ with $B^+ \to K^0 \pi^+$
sets a $1 \sigma$ lower limit of $\gamma > 50^\circ$.  Present $1 \sigma$
bounds on the asymmetry parameter $S_{\pi \pi}$ in $B^0 \to \pi^+ \pi^-$
exclude roughly half the CKM parameter space allowed by other measurements.

We review the flavor decomposition of amplitudes in Section II and the
relevant data in Section III.  The decays $B^+ \to K^0
\pi^+$, expected to be dominated by the penguin amplitude and thus to
have no CP-violating asymmetry, are discussed
in Section IV.  We then analyze rates and CP asymmetries for $B^0
\to K^+ \pi^-$, normalizing amplitudes in terms of the pure-penguin
processes $B^+ \to K^0 \pi^+$, in Section V.  The process
$B^+ \to K^+ \pi^0$ and its comparison with $B^+ \to K^0 \pi^+$
are treated in Section VI, while Section VII deals with $B^0 \to \pi^+ \pi^-$.
Section VIII concludes.

\section{Flavor decomposition of amplitudes}

In order to put the observed rates and asymmetries in theoretical context, we
review the SU(3) flavor-decomposition of $B \to PP$ amplitudes, where $P =
\pi,K$ \cite{GHLR}.  Defining $t = T + P_{EW}^c$, $p = P - \frac{1}{3} P_{EW}^c
- \frac{1}{3}P^E_{EW}$, $c = C + P_{EW}$, $a = A + P^E_{EW}$, and
$e + pa = E + PA + \frac{1}{3}P^A_{EW}$, where $T$ is a color-favored
tree amplitude, $P$ is a penguin amplitude, $C$ is a color-suppressed
tree amplitude, $A$ is an annihilation amplitude, $E$ is an exchange
amplitude, $PA$ is a penguin annihilation amplitude, and $P_{EW}$, $P_{EW}^c$,
$P^E_{EW}$, and $P^A_{EW}$ are respectively color-favored, color-suppressed,
$(\gamma,Z)$-exchange, and $(\gamma,Z)$-direct-channel electroweak penguin
amplitudes \cite{EWVP}, we have
$$
A(B^0 \to \pi^+ \pi^-) = - (t + p + e + pa)~~,~~~
A(B^+ \to \pi^+ \pi^0) = - (t + c)/\s~~~,
$$
$$
A(B^0 \to \pi^0 \pi^0) = (p - c + e + pa)/\sqrt{2}~~,~~~
A(B^0 \to K^0 \ok) = p + pa~~~,
$$
$$
A(B^+ \to \ok K^+) = p + a~~,~~~
A(B^0 \to K^+ K^-) = -(e + pa)~~~.
$$
$$
A(B^0 \to K^+ \pi^-) = -(t' + p')~~,~~~
A(B^+ \to K^0 \pi^+) = p' + a'~~~,
$$
\beq
A(B^+ \to K^+ \pi^0) = -(p' + a' + t' + c')/\s~~,~~~
A(B^0 \to K^0 \pi^0) = (p' - c')/\s~~~,
\eeq
Here unprimed amplitudes denote $\Delta S = 0$ processes, while primed
amplitudes involve $|\Delta S| = 1$.  The $B^0 \to K^+ K^-$ decay is expected
to be highly suppressed since it involves only amplitudes associated with
interactions with the spectator quarks. Measurement
of rates for this process can place upper limits on such spectator
amplitudes (equivalently, on effects of rescattering \cite{resc}).

The quark subprocesses describing the above amplitudes for
$\bar b$ quark decay are summarized in Table \ref{tab:pha}.
We use the unitarity of the CKM matrix, $V^*_{tb}V_{tq} = - V^*_{cb}V_{cq} -
V^*_{ub}V_{uq}, (q=d,s)$, to eliminate elements involving the
top quark in favor of those involving the charm and up quarks in penguin
amplitudes, and then incorporate up quark contributions into redefined tree
contributions. In this convention tree amplitudes involve CKM factors
$V^*_{ub}V_{uq}$, while penguin and electroweak penguin amplitudes contain
factors $V^*_{cb}V_{cq}$. The weak phases of amplitudes for $B$ decays
occur in the last column of Table \ref{tab:pha}.

% This is Table I
\begin{table}
\caption{Weak phases of amplitudes in the flavor decomposition.
\label{tab:pha}}
\begin{center}
\begin{tabular}{c c c c} \hline
Amplitude & Quark & CKM & Weak \\
          & subprocess & element & phase \\ \hline
$T,~C$ & $\bar b \to \bar u u \bar d$ & $V^*_{ub}V_{ud}$ & $\gamma$ \\
$P,~P_{EW},~P_{EW}^c,~P^E_{EW}$ & $\bar b \to \bar d$ & $V^*_{cb}V_{cd}$ &
 0 \\
$E$ & $\bar b d \to \bar u u$ & $V^*_{ub}V_{ud}$ & $\gamma$ \\
$A$ & $\bar b u \to \bar d u$ & $V^*_{ub}V_{ud}$ & $\gamma$ \\
$PA,~P^A_{EW}$ & $\bar b d \to {\rm vacuum}$ & $V^*_{cb}V_{cd}$ & 0 \\
$T',C'$ & $\bar b \to \bar u u \bar s$ & $V^*_{ub}V_{us}$ & $\gamma$ \\
$P',~P'_{EW},~{P'}_{EW}^c,~{P'}^E_{EW}$ & $\bar b \to \bar s$ &
 $V^*_{cb}V_{cs}$ & $\pi$ \\
$E'$ & $\bar b s \to \bar u u$ & $V^*_{ub}V_{us}$ & $\gamma$ \\
%JR                       | |
$A'$ & $\bar b u \to \bar s u$ & $V^*_{ub}V_{us}$ & $\gamma$ \\
$PA',~{P'}^A_{EW}$ & $\bar b s \to {\rm vacuum}$ & $V^*_{cb}V_{cs}$ &
 $\pi$ \\
\hline
\end{tabular}
\end{center}
\end{table}

A useful flavor SU(3) relation between tree and electroweak penguin amplitudes
holds when keeping only dominant $(V-A)(V-A)$ electroweak operators
in the effective weak Hamiltonian. Neglecting very small (a few percent)
electroweak penguin contributions from operators having a different chiral
structure, tree and electroweak penguin operators carrying a given SU(3)
representation are proportional to each other, and one finds in the SU(3)
limit \cite{NR}
\beq\label{tc}
t' + c' = (T' + C')\left (1 - \delta_{EW}e^{-i\gamma}\right )~~,
\eeq
where $\delta_{EW}$ is given in terms of ratios of Wilson coefficients and
CKM factors:
\beq\label{delEW}
\delta_{EW} = -\frac{3}{2}\frac{c_9 + c_{10}}{c_1 + c_2}
\frac{|V^*_{cb}V_{cs}|}{|V^*_{ub}V_{us}|} = 0.65 \pm 0.15~~.
\eeq
The central value is obtained for $|V_{ub}/V_{cb}| = 0.09$.

\section{Rate and asymmetry data and averages}

\leftline{\bf A. Rates}
\medskip

The CLEO \cite{CLrat}, Belle \cite{Berat}, and BaBar \cite{Barat,BaBHF9}
CP-averaged branching ratios for several $B \to PP$ modes are summarized in
Table \ref{tab:rat}, along with averages from Ref.\ \cite{JRTASI}.  We first
note several general properties of these branching ratios.

% This is Table II
\begin{table}
\caption{Branching ratios in units of $10^{-6}$ for $B^0$ or $B^+$ decays to
pairs of light pseudoscalar mesons.  Averages over decay modes and their
CP-conjugates are implied. \label{tab:rat}}
\begin{center}
\begin{tabular}{c c c c c} \hline
Mode & CLEO \cite{CLrat} & Belle \cite{Berat} & BaBar \cite{Barat,BaBHF9} &
Average \cite{JRTASI} \\ \hline
$\pi^+ \pi^-$ & $4.3^{+1.6}_{-1.4} \pm 0.5$ & $5.6^{+2.3}_{-2.0} \pm 0.4$ &
 $4.1 \pm 1.0 \pm 0.7$ & $4.4 \pm 0.9$ \\
$\pi^+ \pi^0$ & $5.4 \pm 2.6$ & $7.8^{+3.8+0.8}_{-3.2-1.2}$ &
 $5.1^{+2.0}_{-1.8} \pm 0.8$ & $5.6 \pm 1.5$ \\
$K^+ \pi^-$ & $17.2^{+2.5}_{-2.4} \pm 1.2$ & $19.3^{+3.4+1.5}_{-3.2-0.6}$ &
 $16.7 \pm 1.6 \pm 1.3$ & $17.4 \pm 1.5$ \\
$K^0 \pi^+$ & $18.2^{+4.6}_{-4.0} \pm 1.6$ & $13.7^{+5.7+1.9}_{-4.8-1.8}$ &
 $18.2^{+3.3}_{-3.0} \pm 2.0$ & $17.3 \pm 2.4$ \\
$K^+ \pi^0$ & $11.6^{+3.0+1.4}_{-2.7-1.3}$ & $16.3^{+3.5+1.6}_{-3.3-1.8}$ &
 $10.8^{+2.1}_{-1.9} \pm 1.0$ & $12.2 \pm 1.7$ \\
$K^0 \pi^0$ & $14.6^{+5.9+2.4}_{-5.1-3.3}$ & $16.0^{+7.2+2.5}_{-5.9-2.7}$ &
 $8.2^{+3.1}_{-2.2} \pm 1.2$ & $10.4 \pm 2.6$ \\
$\pi^0 \pi^0$ & $< 5.6 (90\% {\rm~c.l.})$ &               &            &  \\
$K^0 \ok$ & $< 6.1 (90\% {\rm~c.l.})$ & & $< 7.3  (90\% {\rm~c.l.})$ & \\
$K^+ K^-$ & $< 1.9 (90\% {\rm~c.l.})$ & $< 2.7 (90\% {\rm~c.l.})$ &
 $< 2.5 (90\% {\rm~c.l.})$ & \\
$\ok K^+$ & $< 5.1 (90\% {\rm~c.l.})$ & $< 5.0 (90\% {\rm~c.l.})$ &
 $< 2.4 (90\% {\rm~c.l.})$ & \\
\hline
\end{tabular}
\end{center}
\end{table}

\begin{enumerate}

\item Dominance of $B^0 \to \pi^+ \pi^-$ and $B^+ \to \pi^+ \pi^0$ by the
color-favored tree amplitude would imply the relation
\beq \label{eqn:pirat}
\frac{2 \b(B^+ \to \pi^+ \pi^0)}{\rt \b(B^0 \to \pi^+ \pi^-)} = 1~~~,
\eeq
where $\rt \equiv \tau_{B^+}/\tau_{B^0} = 1.068\pm 0.016$ is the ratio of $B^+$
and $B^0$ lifetimes \cite{Blifes}.  The observed ratio corresponding to the
left-hand side of (\ref{eqn:pirat}) is $2.4 \pm 0.8$, or $1.7 \sigma$ above 1.
The color-suppressed tree amplitude $c$
with Re$(c/t) \simeq 0.2$ \cite{GHLR,BBNS} adds
about 44\% to the predicted $B^+ \to \pi^+ \pi^0$ branching ratio,
converting the right-hand side of (\ref{eqn:pirat}) to 1.44 and reducing
the discrepancy to $1.2 \sigma$.

\item Dominance of the $B \to K \pi$ decays by penguin amplitudes would imply
$$
\b(B^0 \to K^+ \pi^-) = \b(B^+ \to K^0 \pi^+)/\rt
$$
\beq
 = 2 \b(B^+ \to K^+ \pi^0)/\rt = 2 \b(B^0 \to K^0 \pi^0)~~~,
\eeq
while these quantities are in the ratio
\beq
1.08 \pm 0.18~:~1~({\rm def.})~:~1.41 \pm 0.28~:~1.29 \pm 0.37
\eeq
(normalizing to the pure-penguin amplitude for $B^+ \to K^0 \pi^+$).
Thus the strongest evidence for amplitudes other than the penguin
appears at the $1.46 \sigma$ level in the ratio
\beq
R_c \equiv \frac{2 \b(B^+ \to K^+ \pi^0)}{\b(B^+ \to K^0 \pi^+)}
= 1.41 \pm 0.28~~~.
\eeq

\item To first order in subleading amplitudes, one has the sum rule
\cite{HJLsum,GRcomb,JM}
$$
2 \b(B^+ \to K^+ \pi^0)/\rt + 2 \b(B^0 \to K^0 \pi^0)
$$
\beq
= \b(B^+ \to K^0 \pi^+)/\rt + \b(B^0 \to K^+ \pi^-)~~~.
\eeq
The left- and right-hand sides of this relation are $(43.6 \pm 6.1) \times
10^{-6}$ and $(33.6 \pm 2.7) \times 10^{-6}$, respectively. These relations are
fairly general, so any violation of them would most likely signal systematic
experimental errors.

\end{enumerate}

\leftline{\bf B. Asymmetries}
\medskip

In Table \ref{tab:asy} we summarize data on CP asymmetries in $B \to PP$,
defined by
\beq
{\cal A}_{CP} \equiv \frac{\Gamma(\bar B \to \bar f) - \Gamma(B \to f)}
 {\Gamma(\bar B \to \bar f) + \Gamma(B \to f)}~~~,
\eeq
while coefficients of $\sin \Delta m_d t$ and $\cos \Delta m_d t$ measured in
time-dependent CP asymmetries of $\pi^+ \pi^-$ states produced in asymmetric
$e^+ e^-$ collisions at the $\Upsilon(4S)$ are \cite{Gr}
\beq\label{CSpipi}
S_{\pi \pi} \equiv \frac{2 {\rm Im}(\lambda_{\pi \pi})}{1 + |\lambda_{\pi \pi}|
^2}~~,~~~ C_{\pi \pi} \equiv \frac{1 - |\lambda_{\pi \pi}|^2}{1 +
|\lambda_{\pi \pi}|^2}~~~,
\eeq
where
\beq
\lambda_{\pi \pi} \equiv e^{-2i \beta} \frac{A(\ob \to \pi^+ \pi^-)}
{A(B^0 \to \pi^+ \pi^-)}~~~.
\eeq

% This is Table III
\begin{table}
\caption{Asymmetries ${\cal A}_{CP}$ for $B \to PP$ decays
\label{tab:asy}}
\begin{center}
\begin{tabular}{c c c c c} \hline
Mode & CLEO \cite{CLEOasy} & BaBar \cite{BaBasy1,BaBasy2} & Belle \cite{Belasy}
& Average \\ \hline
$S_{\pi \pi}$ & & $0.03^{+0.53}_{-0.56} \pm 0.11$ & & $0.03 \pm 0.56$ \\
$C_{\pi \pi}$ & & $-0.25^{+0.45}_{-0.47} \pm 0.14$ & & $-0.25 \pm 0.48$ \\
$K^+ \pi^-$ & $-0.04 \pm 0.16$ & $-0.07 \pm 0.08 \pm 0.02$ &
 $0.044^{+0.186+0.018}_{-0.167 - 0.021}$ & $-0.048 \pm 0.068$ \\
$K^+ \pi^0$ & $-0.29 \pm 0.23$ & $0.00 \pm 0.18 \pm 0.04$ &
 $-0.059^{+0.222+0.055}_{-0.196-0.017}$ & $-0.096 \pm 0.119$ \\
$K^0 \pi^+$ & $0.18 \pm 0.24$ & $-0.21 \pm 0.18 \pm 0.03$ &
 $0.098^{+0.430+0.020}_{-0.343-0.063}$ & $-0.047 \pm 0.136$ \\ \hline
\end{tabular}
\end{center}
\end{table}

The smallness of these asymmetries will lead to useful constraints on CKM
parameters, though reduction of statistical errors will be quite helpful.
In some cases, however, the reduction of statistical errors on {\it ratios
of branching ratios} described in the previous Section will actually be of
greater use.

\section{$B^+ \to K^0 \pi^+$}

The decay $B^+ \to K^0 \pi^+$ is expected to be dominated by the penguin
amplitude, with a small contribution from the quark subprocess $\bar b u
\to \bar s u$ proportional to the ratio $f_B/m_B \simeq 1/25$.  An equivalent
contribution is generated by rescattering, e.g., from such final states as
$K^+ \pi^0$.  Since the weak phase of the annihilation and penguin amplitudes
are different, the annihilation amplitude can lead to a small CP asymmetry
in the rate for $B^+ \to K^0 \pi^+$ vs.\ its CP-conjugate decay.  There is no
evidence for such an asymmetry at present, but the experimental upper bounds
are no stronger than for processes in which the penguin amplitude is expected
to be accompanied by tree amplitudes, such as $B^0 \to K^+ \pi^-$ and
$B^+ \to K^+ \pi^0$.  Much larger CP asymmetries could occur in those processes
if strong phases were sufficiently large.

A useful way to estimate the effect of the annihilation amplitude in $B^+ \to
K^0 \pi^+$ \cite{Usp} is to use the U-spin \cite{LLM,UMG} transformation
$s \leftrightarrow d$ to relate it to $B^+ \to \ok K^+$.
Under this substitution the penguin amplitude (proportional to
$V^*_{cb}V_{cd}$)  is reduced by a factor of
$\lambda = |V_{cd}/V_{cs}|$, while the annihilation amplitude is {\it
increased} by a factor $\lambda^{-1} = |V_{ud}/V_{us}|$, where $\lambda \simeq
0.22$.  Thus, not only should the CP asymmetry in $B^+ \to \ok K^+$ be
substantially larger than that in $B^+ \to K^0 \pi^+$, but if the annihilation
amplitude is large enough it could lead to an enhancement of the rate for
$B^+ \to \ok K^+$ over that expected if the penguin amplitude $P$ were
dominant, which corresponds to a branching ratio of about $\b(B^+ \to
\ok K^+) \simeq |V_{cd}/V_{cs}|^2 \b(B^+ \to K^0\pi^+) = 8 \times 10^{-7}$
\cite{DGR,CL}.
%JR
The present experimental limit \cite{Barat} is only a factor of three larger.

Evidence for rescattering \cite{resc} would also be forthcoming from the
process $B^0 \to K^+ K^-$, for which the contributions of the $E$ and $PA$
amplitudes are expected to lead to a branching ratio below $10^{-7}$.
Present experimental limits are an order of magnitude above this value.

\section{$B^0 \to K^+ \pi^-$}

Fleischer and Mannel \cite{FM} pointed out that a useful ratio giving
information on the weak phase $\gamma$ is
\beq
R \equiv \frac{\rt \left [\b(\ob \to K^- \pi^+) + \b(B^0 \to K^+ \pi^-)
\right ]} {\b(B^- \to \ok \pi^-) + \b(B^+ \to K^0 \pi^+)}~~~.
\eeq
Within the assumption of a dominant penguin amplitude and a subdominant tree
amplitude, one finds
\beq \label{eqn:R}
R = 1 - 2 r \cos \gamma \cos \delta_0 + r^2~~~,
\eeq
where $r \equiv |T'/P'|$ is the ratio of tree to penguin amplitudes for
strangeness-changing $B$ decays to charmless final states, and $\delta_0
\equiv \delta_{T'} - \delta_{P'}$ is the strong final-state phase difference
between tree and penguin amplitudes.  Independently of $r$ and $\delta_0$
it can then be shown \cite{FM} that $R \ge \sin^2 \gamma$, so that a value of
$R$ below 1 could place useful bounds on $\gamma$.

The present experimental data summarized in Table \ref{tab:rat} indicate
$R = 1.08 \pm 0.18$, so that no useful bound arises from the Fleischer-Mannel
procedure.  However, it was shown in Ref.\ \cite{GRKpi} that if one combined
data on $R$ with the CP pseudo-asymmetry
$$
A_0 \equiv \frac{\Gamma(\ob \to K^- \pi^+) - \Gamma(B^0 \to K^+ \pi^-)}
{\Gamma(B^- \to \ok \pi^-) + \Gamma(B^+ \to K^0 \pi^+)}
= {\cal A}_{CP}(B^0 \to K^+ \pi^-) R
$$
\beq
= -2 r \sin \gamma \sin \delta_0~~~,
\eeq
one could eliminate the strong phase difference between tree and penguin
amplitudes and obtain useful information on the weak phase $\gamma$.
The result is
\beq
R = 1 + r^2 \pm \sqrt{4 r^2 \cos^2 \gamma - A_0^2 \cot^2 \gamma}~~~.
\eeq
Plots of $R$ as a function of $\gamma$ for various values of $r$ and $A_0$
were given in Ref.\ \cite{GRKpi}.  Note that this function is invariant
under the replacement $\gamma \to \pi - \gamma$, so it only need be plotted
for $0 \le \gamma \le 90^\circ$.  However, the expression (\ref{eqn:R})
indicates that the upper branches of the curves correspond to $\cos \gamma
\cos \delta_0 < 0$, while the lower branches correspond to $\cos \gamma \cos
\delta_0 > 0$.

Using the experimental asymmetries summarized in Table \ref{tab:asy}, one
finds $A_0 = -0.052 \pm 0.073$.  In Ref.\ \cite{GRKpi} we estimated $r = 0.16
\pm 0.06$. Using the most recent experimental data for $B^+ \to K^0 \pi^+$
to estimate $|P'|$ and factorization in $B \to \pi l \nu$ \cite{LRfact}
and flavor SU(3) \cite{GHLR} to estimate $|T'|$, an updated result is
$r = 0.184 \pm 0.044$.

% This is Figure 1
\begin{figure}[t]
\centerline{\epsfysize = 5 in \epsffile{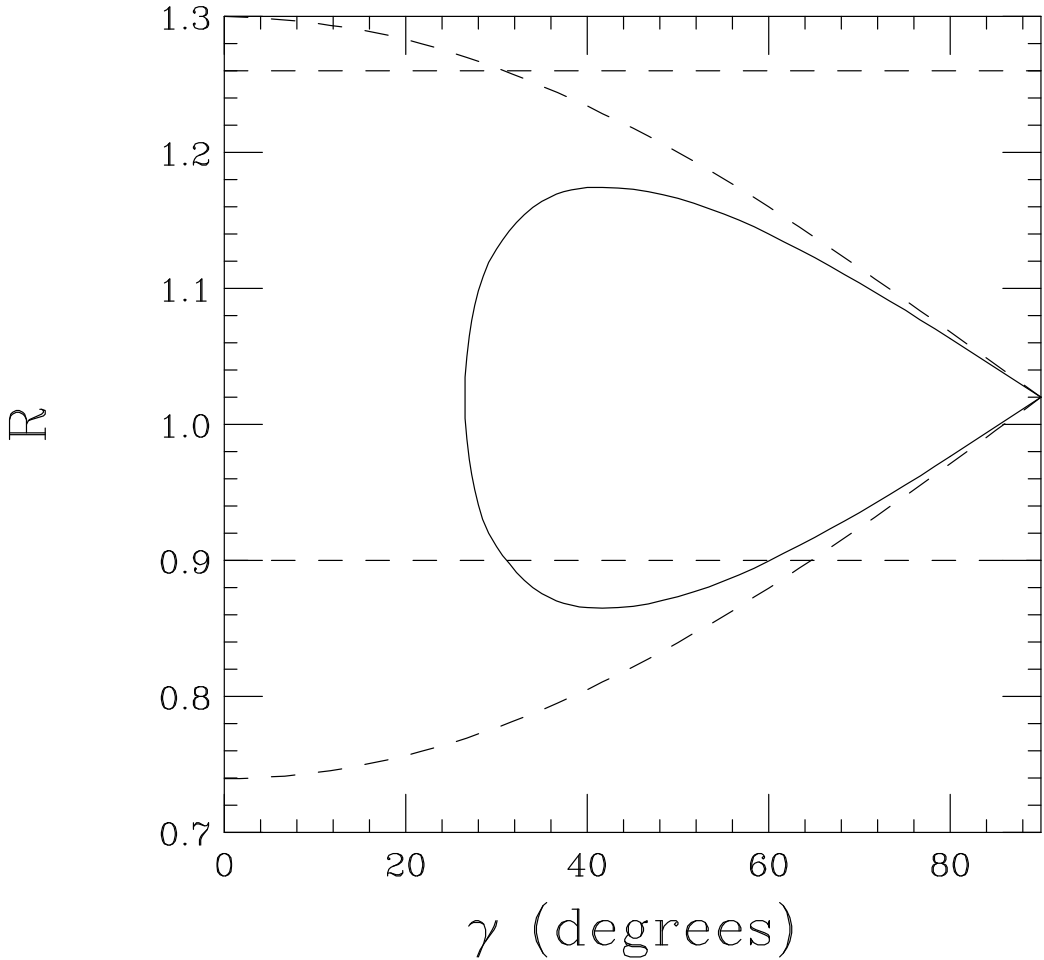}}
\caption{Behavior of $R$ for $r = 0.14$ and $A_0 = 0$ (dashed curves) or
$|A_0| = 0.125$ (solid curve) as a function of the weak phase $\gamma$.
Horizontal dashed lines denote $\pm 1 \sigma$ experimental limits on $R$.
The upper branches of the curves correspond to the case $\cos \gamma
\cos \delta_0 <0$, while the lower branches correspond to $\cos \gamma
\cos \delta_0 >0$.
\label{fig:R}}
\end{figure}

The most conservative bounds on $\gamma$ are obtained using the smallest
value of $r$ and the largest value of $|A_0|$.  A plot of $R$ for $r = 0.14$
(the $1 \sigma$ lower bound) and both $A_0 = 0$ and $|A_0| = 0.125$ (the
$1 \sigma$ upper bound) is shown in Figure \ref{fig:R}.  With present
experimental errors, no useful bound on $\gamma$ emerges from the consideration
of $R$ unless additional assumptions are made.  Reduction of errors on $R$ by
roughly a factor of two could have a considerable impact even given
present errors on $A_0$ and $r$.  Since the curves for $A_0 = 0$ and
$|A_0| = 0.125$ are fairly close to one another for a considerable range
of $\gamma$, improvement of bounds on $A_0$ is less likely to sharpen the
bounds on $\gamma$ unless that angle differs considerably from $90^\circ$.

Theoretical estimates \cite{BBNS} of small final-state phases imply
$\cos \delta_0 > 0$, so that with $\gamma \le 90^\circ$ one
should have destructive tree-penguin interference in $B^0 \to K^+ \pi^-$ and
thus should be on the lower branch of the curves in Fig.\ \ref{fig:R}.  The
$1 \sigma$ lower bound on $R$ then would exclude
$31^\circ \le \gamma \le 60^\circ$.

The expressions for $R$ and $A_0$ are invariant under the interchange of
$\gamma$ and $\delta_0$, so that Fig.\ \ref{fig:R} can also be used in
principle for bounds on $\delta_0$.  At present, no useful bounds emerge.
However, writing $\sin \delta_0 = - A_0/(2 r \sin \gamma)$ and using the
$1 \sigma$ range $-0.125 \le A_0 \le 0.021$ and the lower bounds
$r \ge 0.14$ from the above discussion and $\gamma \ge 32^\circ$ from a fit
to CKM parameters \cite{JRStA}, one finds $- 8^\circ \le \delta_0
\le 57^\circ$
up to a discrete ambiguity which also permits a solution
$\delta_0 \to \pi - \delta_0$.

\section{$B^+ \to K^+ \pi^0$}

The ratio
\beq
R_c \equiv \frac{2[\b(B^- \to K^- \pi^0) + \b(B^+ \to K^+ \pi^0)]}
{\b(B^- \to \ok \pi^-) + \b(B^+ \to K^0 \pi^+)}
\eeq
also contains useful information on the weak phase $\gamma$.  Initially
it was proposed to use this ratio in an amplitude triangle construction
\cite{GLRKpi} in which the amplitude $t'+c' = - A(B^+ \to K^0 \pi^+)
- \s A(B^+ \to K^+ \pi^0)$ was evaluated using flavor SU(3) from the
corresponding amplitude $t + c = -\s A(B^+ \to \pi^+ \pi^0)$.  However,
this procedure neglected important electroweak penguin (EWP) contributions
\cite{EWP}.  It was then shown that these could be taken into account
\cite{NR} through the SU(3) relation (\ref{tc}).
Neglecting $a'$ contributions in decay amplitudes, and writing
\beq
-\s A(B^+\to K^+\pi^0) = p' + (T' + C')\left (1 -
\delta_{EW}e^{-i\gamma}\right )~~,
\eeq
one finds
\beq
R_c = 1 - 2 r_c \cos \delta_c (\cos \gamma - \delta_{EW}) + r_c^2
(1 - 2 \delta_{EW} \cos \gamma + \delta_{EW}^2)~~~,
\eeq
where $r_c \equiv |T'+C'|/|p'|$, $\delta_c \equiv \delta_{T'+C'}
- \delta_{p'}$,
and $\delta_{EW}$ is given in Eq.~(\ref{delEW}). Note that the latter
parameter involves a sizable uncertainty from $|V_{ub}/V_{cb}|$.
In order to demonstrate possible constraints on weak and strong phases, we will
explore the effect of $\pm 1 \sigma$ deviations from the central value of
$\delta_{EW}=0.65 \pm 0.15$.

The CP-violating asymmetry in $B^+ \to K^+ \pi^0$ decays
then provides a constraint on the relative strong phase $\delta_c$.  We define
a pseudo-asymmetry
$$
A_c \equiv \frac{2[\b(B^- \to K^- \pi^0) - \b(B^+ \to K^+ \pi^0)]}
{\b(B^- \to \ok \pi^-) + \b(B^+ \to K^0 \pi^+)}
= R_c {\cal A}_{CP}(B^+ \to K^+ \pi^0)
$$
\beq
= - 2 r_c \sin \delta_c \sin \gamma~~~,
\eeq
and, using the experimental averages in Tables \ref{tab:rat} and
\ref{tab:asy}, we find $A_c = -0.13 \pm 0.17$.

\begin{figure}
\centerline{\epsfysize = 4.9 in \epsffile{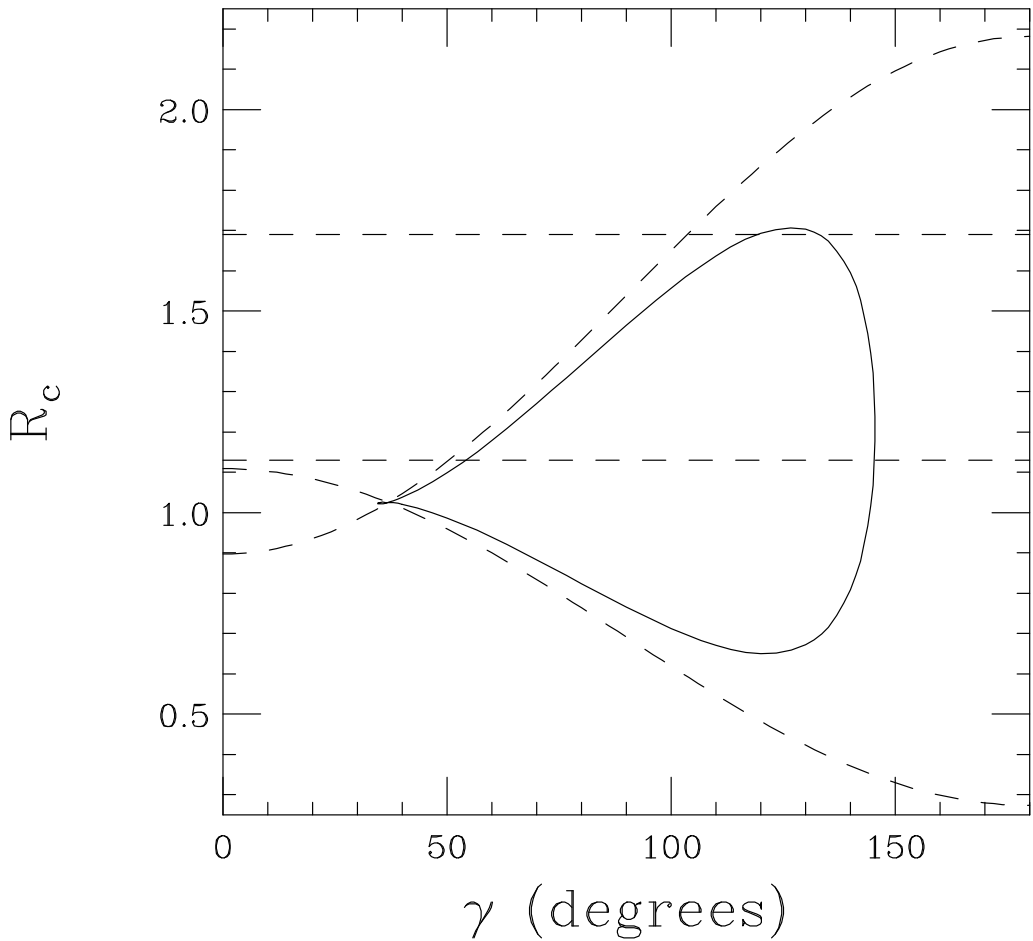}}
\caption{Behavior of $R_c$ for $r_c = 0.265$ ($1 \sigma$ upper limit) and
$A_c = 0$ (dashed curves) or $|A_c| = 0.30$ (solid curve) as a function of the
weak phase $\gamma$. Horizontal dashed lines denote $\pm 1 \sigma$ experimental
limits on $R_c$.  Upper branches of curves correspond to
$\cos \delta_c(\cos \gamma - \delta_{EW}) < 0$, while lower branches
correspond to $\cos \delta_c(\cos \gamma - \delta_{EW}) > 0$.  Here we have
taken $\delta_{EW} = 0.80$ [the $1 \sigma$ upper limit in Eq.\ (3)], which
leads to the most conservative bound on $\gamma$.
\label{fig:Rc}}
\end{figure}

Eliminating $\delta_c$, we can plot $R_c$ as a function of $\gamma$ for
various values of $\delta_{EW}$, $r_c$ and $A_c$, to see if any constraints on
$\gamma$ emerge when taking a $1\sigma$ lower limit on $R_c$, $R_c \ge 1.13$.
The ratio $r_c$, obtained from \cite{GLRKpi}
\beq
r_c = \s \frac{V_{us}}{V_{ud}}\frac{f_K}{f_\pi}\left[\frac
{\b(B^- \to \pi^- \pi^0) + \b(B^+ \to \pi^+ \pi^0)}
{\b(B^- \to \ok \pi^-) + \b(B^+ \to K^0 \pi^+)}\right]^{1/2}~~,
\eeq
was estimated in Ref.\ \cite{NR} to be $r_c \equiv \epsilon_{3/2} = 0.24 \pm
0.06$.  We can update this estimate using the new branching
ratios quoted in Table \ref{tab:rat}, finding $r_c = 0.230 \pm 0.035$.
The resulting plot is shown in Fig.\ \ref{fig:Rc} for the $+1 \sigma$ values
of $r_c$ and $\delta_{EW}$ (which lead to the weakest lower bound on $\gamma$),
both for $A_c = 0$ and for the $1 \sigma$ upper limit $A_c = 0.30$.
The weakest $1 \sigma$ bound on $\gamma$ in this case, as opposed to the
case of $B^0 \to K^+ \pi^-$, occurs when $A_c = 0$, and
is $\gamma \ge 50^\circ$.  As a result of the electroweak penguin term, the
value of $R_c$ is not symmetric under the replacement $\gamma \to \pi -
\gamma$, in contrast to the case of $R$ for $B^0 \to K^+ \pi^-$.  In Table
\ref{tab:gammas} we show the minimum values of $\gamma$ obtained on the basis
of the $1 \sigma$ inequality $R_c \ge 1.13$ for $r_c = 0.230 \pm 0.035$ and
$\delta_{EW} = 0.65 \pm 0.15$, both for $A_c = 0$ and for $A_c = 0.3$.

% This is Figure 2

% This is Table IV
\begin{table}
\caption{Minimum values of $\gamma$ (in degrees)
for $R_c \ge 1.13$, given central and $\pm
1 \sigma$ values of $r_c$ and $\delta_{EW}$.  First figure denotes value with
$A_c = 0$ while second figure denotes value with $|A_c| = 0.30$.
\label{tab:gammas}}
\begin{center}
\begin{tabular}{c c c c} \hline
\qquad $r_c$: & 0.195 & 0.230 & 0.265 \\
$\delta_{EW}$: &      &       &       \\ \hline
0.50 & 75/82 & 71/74 & 68/70 \\
0.65 & 66/74 & 62/67 & 60/62 \\
0.80 & 57/68 & 53/59 & 50/54 \\ \hline
\end{tabular}
\end{center}
\end{table}

As in the case of $B^0 \to K^+ \pi^-$, there is little difference on the
bounds one obtains for zero CP asymmetry and for the maximum allowed value.
The greatest leverage on bounds would be provided by reducing the experimental
error on $R_c$, with some additional help associated with reduction of the
errors on $r_c$ and $\delta_{EW}$.
The limits of Table \ref{tab:gammas} correspond to the
branches of the curves that would be chosen if $\cos \delta > 0$, as
expected in some theoretical treatments \cite{BBNS}.

One can place a one-sided $1 \sigma$ limit on the strong phase $\delta_c$ using
the present range $-0.30 \le A_c \le 0.04$.  With
\beq
\sin \delta_c = - A_c/(2 r_c \sin \gamma)~~~,
\eeq
$\gamma \ge 32^\circ$, and $r_c \ge 0.195$ one has $-0.19 \le \sin \delta_c \le
1.44$, so $\delta_c \ge -11^\circ$.  The upper limit on $|A_c|$ [equivalently,
on $|{\cal A}_{CP}(B^+ \to K^+ \pi^0)$] must be reduced to about 2/3 of its
present value if a two-sided constraint on $\delta_c$ is to be obtained.

\section{$B^0 \to \pi^+ \pi^-$}

The implications of the BaBar \cite{BaBasy2} limits on $S_{\pi \pi}$ and
$C_{\pi \pi}$ quoted in Table \ref{tab:asy} have been partially explored
in Ref.\ \cite{LRfact}.  Here we review these limits, discuss their
implications for CKM parameters, and discuss prospects for their improvement.

As mentioned in Section III, the present experimental ratio (\ref{eqn:pirat})
of $B^+ \to \pi^+ \pi^0$ and $B^0 \to \pi^+ \pi^-$ branching ratios is
somewhat larger than that expected from tree-dominance alone, even
accounting for a color-suppressed contribution to the former process.  For
this reason, as well as for the purpose of estimating the ``penguin pollution"
correction to the time-dependent CP asymmetry in $B^0 \to \pi^+ \pi^-$, it is
useful to estimate the ratio $|P/T|$ of penguin to tree amplitudes in $\Delta S
= 0$ $B$ decays.  Using this estimate it is then possible to place limits
on the weak phase $\alpha$ even given the crude limits on $S_{\pi \pi}$ and
$C_{\pi \pi}$ noted in Table \ref{tab:asy}.

Many previous attempts have been made to estimate $|P/T|$ in a
model-independent way, including an isospin analysis requiring the measurement
of $B^+ \to \pi^+ \pi^0$, $B^0 \to \pi^0 \pi^0$,
and corresponding charge-conjugate decays \cite{GL}, methods
which use only part of the above information \cite{GQ,Ch,GLSS}, and numerous
applications of flavor SU(3) \cite{GHLR,SilWo,SU}.  There have been hints,
based on earlier data, that the penguin amplitude was interfering
destructively with the tree in $B^0 \to \pi^+ \pi^-$ \cite{dest}.

The method of Ref.\ \cite{LRfact} is capable in principle of giving a good
value of $|T|$ based on na\"{\i}ve factorization and measurement of the
spectrum of $B \to \pi l \nu$ near $q^2 = 0$, where $q^2$ is the squared
effective mass of the $l \nu$ system.  Present experimental measurements
and some theoretical estimates of form factor shapes based on lattice
gauge theory lead to an estimate $|T| = 2.7 \pm 0.6$, where all amplitudes are
quoted as square roots of $\bo$ branching ratios multiplied by $10^3$.  This is
the same value obtained \cite{B2Kfact} from $B^+ \to
\pi^+ \pi^0$ with additional assumptions about the color-suppressed amplitude.

The penguin amplitude can be estimated from $B^+ \to K^0 \pi^+$.  The average
of the branching ratios for that process in Table \ref{tab:rat} is
\beq
{\cal B}(B^+ \to K^0 \pi^+) = (17.2 \pm 2.4) \times 10^{-6}~~~,
\eeq
leading to $|P'|^2 = (17.2 \pm 2.4)/\rt$, $|P'| = 4.02 \pm 0.28$,

We now estimate the strangeness-preserving $\bar b \to \bar d$ amplitude $|P|$
which is proportional to the CKM factor $V_{cd} V^*_{cb}$ in our convention.
We find
\beq
|P/P'| = \left| \frac{V_{cd}}{V_{cs}} \right|
= 0.22~~,~~~|P| \simeq 0.91 \pm 0.06~~~,
\eeq
Assuming factorization of penguin amplitudes \cite{BBNS}, this estimate
is corrected by an SU(3) breaking factor of $f_{\pi}/f_K$ and becomes
$|P| \simeq 0.74 \pm 0.05$.

With the present method of estimating errors on $|P|$ and $|T|$, we then
find $|P/T| = 0.34 \pm 0.08$ without introducing SU(3) breaking in
$P/P'$, or $|P/T| = 0.276 \pm 0.064$ when SU(3) breaking in $P/P'$ is
introduced through $f_\pi/f_K$. The latter number,
%JR
which will be used in the subsequent discussion,
is to be compared with a value of
$0.285 \pm 0.076$ obtained by \cite{BBNS} on the basis of a theoretical
calculation which includes small annihilation corrections. A value of
%JR          | (We re-calculated the error)
$0.26 \pm 0.08$ was obtained \cite{LRfact} when defining $P$ and $P'$
as the amplitudes containing $V_{td}$ and $V_{ts}$, respectively, without
introducing SU(3) breaking in the ratio of these amplitudes.

The decay amplitudes to $\pi^+ \pi^-$ for $B^0$ and $\ob$ are
$$
A(\bo \to \pi^+ \pi^-) = -(|T|e^{i \delta_T} e^{i \gamma} +
 |P| e^{i \delta_P})~~~,
$$
\beq
A(\ob \to \pi^+ \pi^-) = -(|T|e^{i \delta_T} e^{- i \gamma} +
 |P| e^{i \delta_P})~~~,
\eeq
where $\delta_T$ and $\delta_P$ are strong phases of the tree and penguin
amplitudes, and
$\delta \equiv \delta_P - \delta_T$.  The CP-averaged
branching ratio in Table \ref{tab:rat} then implies
\beq \label{eqn:pipirat}
|T|^2 + |P|^2 + 2 |TP| \cos \gamma \cos \delta = 4.4 \pm 0.9~~~,
\eeq
which suggests but does not prove, given our errors on $|T|$ and $|P|$,
that the tree and penguin amplitudes are interfering destructively with one
another in $B^0 \to \pi^+ \pi^-$.  For $\cos \delta >0$ as favored
theoretically \cite{BBNS}, this would require $\cos \gamma < 0$, which is
not favored by CKM fits \cite{JRTASI}.

\begin{figure}[t]
\centerline{\epsfysize = 3 in \epsffile{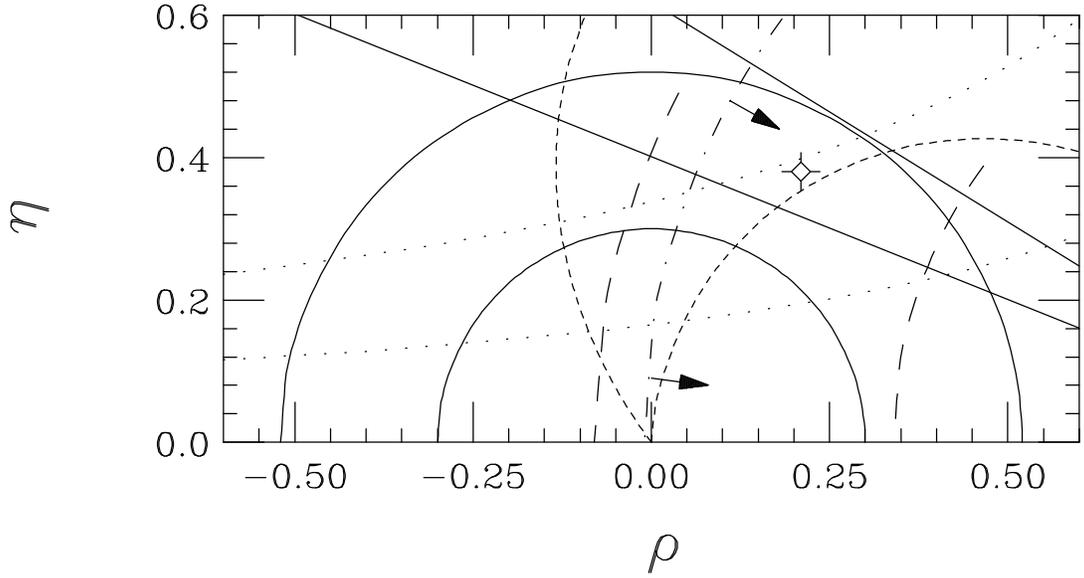}}
\caption{Constraints on parameters of the CKM matrix.  Solid circles denote
limits on $|V_{ub}/V_{cb}| = 0.090 \pm 0.025$ from charmless $b$ decays.
Dashed arcs denote limits from $\bo$--$\ob$ mixing.  Dot-dashed arc denotes
limit from $B_s$--$\overline{B}_s$ mixing.  Dotted hyperbolae are associated
with limits on CP-violating $K^0$--$\ok$ mixing (the parameter $\epsilon$).
Limits of $\pm 1 \sigma$ from CP asymmetries in $B^0 \to J/\psi K_S$
leading to $\sin(2 \beta) = 0.79 \pm 0.10$ are shown by the solid rays.
The small dashed lines represent the constraint due to $S_{\pi \pi}$,
with $0.21 \le |P/T| \le 0.34$.
The plotted point lies in the middle of the allowed region.
\label{fig:re}}
\end{figure}

The BaBar Collaboration \cite{BaBasy2} has recently reported the first results
for the CP-violating asymmetries (\ref{CSpipi}) in $B^0 \to \pi^+ \pi^-$
decays.  Our expressions for the decay amplitudes imply
\beq
\lambda_{\pi \pi}=e^{2 i \alpha} \left( \frac{1 + |P/T| e^{i \delta}
e^{i \gamma}}{1 + |P/T| e^{i \delta} e^{-i \gamma}} \right)~~~.
\eeq
In the absence of the penguin amplitude we would have $S_{\pi \pi} = \sin(2
\alpha)$.  If $|P/T| \ne 0$ but $\delta$ is small \cite{BBNS}, we
have $S_{\pi \pi} \simeq \sin(2 \alpha_{\rm eff}$),
where $\alpha_{\rm eff} = \alpha + \Delta \alpha$, with
\beq
\Delta \alpha = \tan^{-1} \left[ \frac{|P/T| \sin \gamma}{1 + |P/T| \cos
\gamma} \right]~~~.
\eeq
Using
\beq
\tan \alpha = \frac{\eta}{\eta^2 - \rho(1-\rho)}~~,~~~ \tan{\Delta \alpha}
= \frac{\eta|P/T|}{\sqrt{\rho^2 + \eta^2} + \rho|P/T|}~~~,
\eeq
we plot in Fig.~\ref{fig:re} the $\pm 1 \sigma$ contours of $-0.53 \le S_{\pi
\pi} \le 0.59$, along with other CKM constraints taken from Ref.~\cite{JRTASI}.
The $1\sigma~S_{\pi \pi}$ bounds exclude about half of the $(\rho,\eta)$
parameter space allowed by all other constraints.
Similar constraints under slightly different technical
assumptions were obtained in Ref.\ \cite{LRfact}.

% This is Figure 3

The quantity $C_{\pi \pi}$ is also consistent at present with zero.  Its
observed range is not yet tightly enough constrained to provide much
information, but reduction in errors will eventually be useful mainly
in constraining the strong phase difference $\delta$.  For
one such example, see Ref.\ \cite{LRfact}.

\section{Conclusions}

While a CP-violating indirect asymmetry (associated with $B^0$--$\ob$ mixing)
has been observed in the decays $B^0 \to J/\psi K_S$, no direct asymmetries
have yet been observed in $B \to K \pi$ decays, and no asymmetries of any
sort have been seen in $B \to \pi^+ \pi^-$.  Nonetheless, the present upper
limits on $K \pi$ and $\pi \pi$ asymmetries, crude as they are, already are
beginning to provide useful information on CKM phases.
As one example, the deviation of the ratio $2 \b(B^+ \to K^+ \pi^0)/\b(B^+
\to K^0 \pi^+)$ from 1 is able at the $1 \sigma$ level to provide a lower
bound $\gamma \ge 50^\circ$
{\it independently} of the CP asymmetry in $B^+ \to K^+ \pi^0$.
The proximity of the ratio $(\tau_+/\tau_0)
\b(B^0 \to K^+ \pi^-)/\b(B^+ \to K^0 \pi^+)$ to unity, when combined with
the expectation that the final-state strong phase is small in the $K^+ \pi^-$
system, allows one to exclude a range $31^\circ \le \gamma \le 60^\circ$ at the
$1 \sigma$ level.  Finally, the $\pm 1 \sigma$ bounds on $S_{\pi \pi}$ allow
one to exclude (at the $1 \sigma$ level) roughly half of the parameter
space in the $(\rho,\eta)$ plane allowed by other observables.  The $1 \sigma$
bound $\gamma \ge 60^\circ$ is the strongest constraint of these.

Uncertainties in theoretical parameters, including the ratios of tree to
penguin amplitudes in $B\to K\pi$ and $B\to \pi\pi$, should be reduced
in the future with a larger amount of data. More severe constraints
are expected for small rescattering and color-suppressed electroweak
%JR
amplitudes in $B \to K \pi$, which were neglected in the present treatment.
With the increased data samples expected to be available from BaBar and Belle,
one can look forward to greatly improved limits on CKM parameters from
analyses such as ours even if no CP asymmetries are observed in $B \to K \pi$
and $B \to \pi \pi$ decays.

\section*{Acknowledgments}

We thank Martin Beneke, Alex Kagan and Vivek Sharma for discussions.
This work was performed in part at the Aspen Center for Physics.  The work
of J. L. R. was supported in part by the United States Department of Energy
through Grant No.\ DE FG02 90ER40560. This work was partially supported by
the Israel Science Foundation founded by the Israel Academy of Sciences and
Humanities and by the US - Israel Binational Science Foundation through
Grant No. 98-00237.

% Journal definitions
\def \ajp#1#2#3{Am.\ J. Phys.\ {\bf#1}, #2 (#3)}
\def \apny#1#2#3{Ann.\ Phys.\ (N.Y.) {\bf#1}, #2 (#3)}
\def \app#1#2#3{Acta Phys.\ Polonica {\bf#1}, #2 (#3)}
\def \arnps#1#2#3{Ann.\ Rev.\ Nucl.\ Part.\ Sci.\ {\bf#1}, #2 (#3)}
\def \art{and references therein}
\def \cmts#1#2#3{Comments on Nucl.\ Part.\ Phys.\ {\bf#1}, #2 (#3)}
\def \cn{Collaboration}
\def \cp89{{\it CP Violation,} edited by C. Jarlskog (World Scientific,
Singapore, 1989)}
\def \econf#1#2#3{Electronic Conference Proceedings {\bf#1}, #2 (#3)}
\def \efi{Enrico Fermi Institute Report No.}
\def \epjc#1#2#3{Eur.\ Phys.\ J.\ C {\bf#1}, #2 (#3)}
\def \f79{{\it Proceedings of the 1979 International Symposium on Lepton and
Photon Interactions at High Energies,} Fermilab, August 23-29, 1979, ed. by
T. B. W. Kirk and H. D. I. Abarbanel (Fermi National Accelerator Laboratory,
Batavia, IL, 1979}
\def \hb87{{\it Proceeding of the 1987 International Symposium on Lepton and
Photon Interactions at High Energies,} Hamburg, 1987, ed. by W. Bartel
and R. R\"uckl (Nucl.\ Phys.\ B, Proc.\ Suppl., vol. 3) (North-Holland,
Amsterdam, 1988)}
\def \ib{{\it ibid.}~}
\def \ibj#1#2#3{~{\bf#1}, #2 (#3)}
\def \ichep72{{\it Proceedings of the XVI International Conference on High
Energy Physics}, Chicago and Batavia, Illinois, Sept. 6 -- 13, 1972,
edited by J. D. Jackson, A. Roberts, and R. Donaldson (Fermilab, Batavia,
IL, 1972)}
\def \ijmpa#1#2#3{Int.\ J.\ Mod.\ Phys.\ A {\bf#1}, #2 (#3)}
\def \ite{{\it et al.}}
\def \jhep#1#2#3{JHEP {\bf#1}, #2 (#3)}
\def \jpb#1#2#3{J.\ Phys.\ B {\bf#1}, #2 (#3)}
\def \lg{{\it Proceedings of the XIXth International Symposium on
Lepton and Photon Interactions,} Stanford, California, August 9--14, 1999,
edited by J. Jaros and M. Peskin (World Scientific, Singapore, 2000)}
\def \lkl87{{\it Selected Topics in Electroweak Interactions} (Proceedings of
the Second Lake Louise Institute on New Frontiers in Particle Physics, 15 --
21 February, 1987), edited by J. M. Cameron \ite~(World Scientific, Singapore,
1987)}
\def \kaon{{\it Kaon Physics}, edited by J. L. Rosner and B. Winstein,
%U                              |
University of Chicago Press, 2001}
\def \kdvs#1#2#3{{Kong.\ Danske Vid.\ Selsk., Matt-fys.\ Medd.} {\bf #1}, No.\
#2 (#3)}
\def \ky{{\it Proceedings of the International Symposium on Lepton and
Photon Interactions at High Energy,} Kyoto, Aug.~19-24, 1985, edited by M.
Konuma and K. Takahashi (Kyoto Univ., Kyoto, 1985)}
\def \mpla#1#2#3{Mod.\ Phys.\ Lett.\ A {\bf#1}, #2 (#3)}
\def \nat#1#2#3{Nature {\bf#1}, #2 (#3)}
\def \nc#1#2#3{Nuovo Cim.\ {\bf#1}, #2 (#3)}
\def \nima#1#2#3{Nucl.\ Instr.\ Meth.\ A {\bf#1}, #2 (#3)}
\def \np#1#2#3{Nucl.\ Phys.\ {\bf#1}, #2 (#3)}
\def \npps#1#2#3{Nucl.\ Phys.\ Proc.\ Suppl.\ {\bf#1}, #2 (#3)}
\def \os{XXX International Conference on High Energy Physics, Osaka, Japan,
July 27 -- August 2, 2000}
\def \PDG{Particle Data Group, D. E. Groom \ite, \epjc{15}{1}{2000}}
\def \pisma#1#2#3#4{Pis'ma Zh.\ Eksp.\ Teor.\ Fiz.\ {\bf#1}, #2 (#3) [JETP
Lett.\ {\bf#1}, #4 (#3)]}
\def \pl#1#2#3{Phys.\ Lett.\ {\bf#1}, #2 (#3)}
\def \pla#1#2#3{Phys.\ Lett.\ A {\bf#1}, #2 (#3)}
\def \plb#1#2#3{Phys.\ Lett.\ B {\bf#1}, #2 (#3)}
\def \pr#1#2#3{Phys.\ Rev.\ {\bf#1}, #2 (#3)}
\def \prc#1#2#3{Phys.\ Rev.\ C {\bf#1}, #2 (#3)}
\def \prd#1#2#3{Phys.\ Rev.\ D {\bf#1}, #2 (#3)}
\def \prl#1#2#3{Phys.\ Rev.\ Lett.\ {\bf#1}, #2 (#3)}
\def \prp#1#2#3{Phys.\ Rep.\ {\bf#1}, #2 (#3)}
\def \ptp#1#2#3{Prog.\ Theor.\ Phys.\ {\bf#1}, #2 (#3)}
\def \rmp#1#2#3{Rev.\ Mod.\ Phys.\ {\bf#1}, #2 (#3)}
\def \rp#1{~~~~~\ldots\ldots{\rm rp~}{#1}~~~~~}
\def \si90{25th International Conference on High Energy Physics, Singapore,
Aug. 2-8, 1990}
\def \slc87{{\it Proceedings of the Salt Lake City Meeting} (Division of
Particles and Fields, American Physical Society, Salt Lake City, Utah, 1987),
ed. by C. DeTar and J. S. Ball (World Scientific, Singapore, 1987)}
\def \slac89{{\it Proceedings of the XIVth International Symposium on
Lepton and Photon Interactions,} Stanford, California, 1989, edited by M.
Riordan (World Scientific, Singapore, 1990)}
\def \smass82{{\it Proceedings of the 1982 DPF Summer Study on Elementary
Particle Physics and Future Facilities}, Snowmass, Colorado, edited by R.
Donaldson, R. Gustafson, and F. Paige (World Scientific, Singapore, 1982)}
\def \smass90{{\it Research Directions for the Decade} (Proceedings of the
1990 Summer Study on High Energy Physics, June 25--July 13, Snowmass,
Colorado),
edited by E. L. Berger (World Scientific, Singapore, 1992)}
\def \tasi{{\it Testing the Standard Model} (Proceedings of the 1990
Theoretical Advanced Study Institute in Elementary Particle Physics, Boulder,
Colorado, 3--27 June, 1990), edited by M. Cveti\v{c} and P. Langacker
(World Scientific, Singapore, 1991)}
\def \TASI{{\it TASI-2000:  Flavor Physics for the Millennium}, edited by J. L.
Rosner (World Scientific, 2001)}
\def \yaf#1#2#3#4{Yad.\ Fiz.\ {\bf#1}, #2 (#3) [Sov.\ J.\ Nucl.\ Phys.\
{\bf #1}, #4 (#3)]}
\def \zhetf#1#2#3#4#5#6{Zh.\ Eksp.\ Teor.\ Fiz.\ {\bf #1}, #2 (#3) [Sov.\
Phys.\ - JETP {\bf #4}, #5 (#6)]}
\def \zpc#1#2#3{Zeit.\ Phys.\ C {\bf#1}, #2 (#3)}
\def \zpd#1#2#3{Zeit.\ Phys.\ D {\bf#1}, #2 (#3)}


\begin{thebibliography}{99}

\bibitem{GRKpi} M. Gronau and J. L. Rosner, \prd{57}{6843}{1998}.

\bibitem{NR} M. Neubert and J. L. Rosner, \plb{441}{403}{1998};
\prl{81}{5076}{1998}; M. Neubert, \jhep{9902}{014}{1999};
M. Gronau, D. Pirjol and T. M. Yan, \prd{60}{034021}{1999}.

\bibitem{GRpipi} M. Gronau and J. L. Rosner, \prl{76}{1200}{1996};
A. S. Dighe, M. Gronau, and J. L. Rosner, \prd{54}{3309}{1996};
A. S. Dighe and J. L. Rosner, \prd{54}{4677}{1996}.

\bibitem{PP} M. Gronau, J. Rosner and D. London, \prl{73}{21}{1994};
R. Fleischer, \plb{365}{399}{1996}; \prd{58}{093001}{1998};
R. Fleischer and T. Mannel, \prd{57}{2752}{1998};
A. J. Buras, R. Fleischer, and T. Mannel, \np{B533}{3}{1998};
R. Fleischer and A. J. Buras, \epjc{11}{93}{1999};
M. Gronau and D. Pirjol, \prd{61}{013005}{2000};
A. J. Buras and R. Flesicher, \epjc{16}{97}{2000};
M. Gronau, \nima{462}{1}{2001};
M. Gronau and J. L. Rosner, \plb{500}{247}{2001};
X.-G. He, Y. K. Hsiao, J. Q. Shi, Y. L. Wu, and Y. F. Zhou, \prd{64}{034002}
{2001}.

\bibitem{CLEOasy} CLEO \cn, S. Chen \ite, \prl{85}{525}{2000}.

% [6] updated
\bibitem{BaBasy1} BaBar \cn, B. Aubert \ite, \prl{87}{151802}{2001}.

\bibitem{BaBasy2} BaBar \cn, B. Aubert \ite, SLAC report SLAC-PUB-8929,
hep-ex/0107074.

% [8] updated
\bibitem{Belasy} Belle \cn, K. Abe \ite, \prd{64}{071101}{2001}.

\bibitem{GHLR} M. Gronau, O. F. Hern\'andez, D. London, and J. L. Rosner,
\prd{50}{4529}{1994}; \prd{52}{6374}{1995}.

\bibitem{EWVP} M. Gronau, \prd{62}{014031}{2000}.

\bibitem{resc} B. Blok, M. Gronau, and J. L. Rosner, \prl{78}{3999}{1997};
\ibj{79}{1167}{1997}; M. Gronau and J. L. Rosner, \prd{58}{113005}{1998}.

\bibitem{CLrat} CLEO \cn, D. Cronin-Hennessy \ite, \prl{85}{515}{2000};
S. J. Richichi \ite, \prl{85}{520}{2000}; D. Cinabro, presented at XXX
International Conference on High Energy Physics, Osaka, Japan, July 27 --
August 2, 2000, hep-ex/0009045.  The CLEO $B^+ \to \pi^+ \pi^0$ branching
ratio is estimated by M. Gronau and J. L. Rosner, \prd{61}{073008}{2000}.

% [13] updated
\bibitem{Berat} Belle \cn, K. Abe \ite, \prl{87}{101801}{2001}

\bibitem{Barat} BaBar \cn, T. Champion, Osaka Conf.\ \cite{CLrat},
hep-ex/0011018; G. Cavoto, XXXVI Rencontres de Moriond, March 17--24,
2001 (unpublished); B. Aubert \ite, \cite{BaBasy1}.

\bibitem{BaBHF9} BaBar \cn, B. Aubert \ite, SLAC report SLAC-PUB-8978,
hep-ex/0109005, submitted to the 9th International Symposium on Heavy Flavor
Physics, Sept.\ 10--13, 2001, Pasadena, CA.

\bibitem{JRTASI} J. L. Rosner, lectures at TASI-2000, Boulder, Colorado,
June 5--30, 2000, \efi~2000-47, hep-ph/0011355, to be published in \TASI.

\bibitem{Blifes} K. Osterberg, talk presented at the International Europhysics
Conference on High-Energy Physics, Budapest, Hungary, 12--18 July 2001,
to appear in the Proceedings.

\bibitem{BBNS} M. Beneke, G. Buchalla, M. Neubert, and C. T. Sachrajda,
\np{B606}{245}{2001}.

\bibitem{HJLsum} H. J. Lipkin, \plb{445}{403}{1999}.

\bibitem{GRcomb} M. Gronau and J. L. Rosner, \prd{59}{113002}{1999}.

\bibitem{JM} J. Matias, Univ.\ of Barcelona report UAB-FT-514,
hep-ph/0105103 (unpublished).

\bibitem{Gr} M. Gronau, \prl{63}{1451}{1989}.

\bibitem{Usp} A. Falk, A. L. Kagan, Y. Nir, and A. Petrov,
\prd{57}{4290}{1998}.

\bibitem{LLM} S. Meshkov, C. A. Levinson, and H. J. Lipkin, \prl{10}{361}
{1963}.

\bibitem{UMG} M. Gronau, \plb{492}{297}{2000}.

\bibitem{DGR} A. S. Dighe, M. Gronau and J. L. Rosner, \prl{79}{4333}{1997}.

\bibitem{CL} A somewhat larger branching ratio was calculated recently
by C. H. Chen and H. N. Li, \prd{63}{014003}{2001}.

\bibitem{FM} R. Fleischer and T. Mannel, \prd{57}{2752}{1998}.

\bibitem{LRfact} Z. Luo and J. L. Rosner, \efi~01-28, hep-ph/0108024,
submitted to Phys.\ Rev.\ D.

\bibitem{JRStA} J. L. Rosner, \efi~01-34, hep-ph/0108195, based on five
lectures at the 55th Scottish Universities' Summer School in Particle Physics,
St.\ Andrews, Scotland, August 7--23, 2001, to be published by the
Institute of Physics (U.K.).

\bibitem{GLRKpi} M. Gronau, D. London, and J. L. Rosner, \prl{73}{21}{1994}.

\bibitem{EWP} R. Fleischer, \zpc{62}{81}{1994}; \plb{321}{259}{1994};
\ibj{332}{419}{1994}; \ibj{365}{399}{1994}; N. G. Deshpande and
X.-G. He, \prl{74}{26, 4099(E)}{1995}.

\bibitem{GL} M. Gronau and D. London, \prl{65}{3381}{1990}.

\bibitem{GQ} Y. Grossman and H. R. Quinn, \prd{58}{017504}{1998}.

\bibitem{Ch} J. Charles, \prd{59}{054007}{1999}.

\bibitem{GLSS} M. Gronau, D. London, N. Sinha, and R. Sinha, \plb{514}{315}
{2001}.

\bibitem{SilWo} J. Silva and L. Wolfenstein, \prd{49}{1151}{1994}.

\bibitem{SU} See, e.g., Ref.\ \cite{GRpipi}.

\bibitem{dest} W.-S. Hou, J. G. Smith, and F. W\"urthwein, hep-ex/9910014;
X.-G. He, W.-S. Hou, and K. C. Yang, \prl{83}{1100}{1999};
M. Gronau and J. L. Rosner, \prd{61}{073008}{2000}.

\bibitem{B2Kfact} J. L. Rosner, \nima{462}{44}{2001}.

\end{thebibliography}
\end{document}